\documentclass[
    ,final            
  ]
  {aipproc}

\layoutstyle{6x9}

\begin{document}

\title{Hadron Spectroscopy in 2006}


\author{Jonathan L. Rosner}{
  address={Enrico Fermi Institute and Department of Physics,
  University of Chicago, 5640 S. Ellis Avenue, Chicago IL 60637 USA}
}

\begin{abstract}
New results on hadron spectra have been appearing in abundance in the past few
years as a result of improved experimental techniques.  These include
information on states made of both light quarks (u, d, and s) and with one or
more heavy quarks (c, b).  The present review, dedicated to the memory of R. H.
Dalitz, treats light-quark states, glueballs, hybrids, charmed and beauty
particles, charmonium, and $b \bar b$ states.  Some future directions are
mentioned.
\end{abstract}

\maketitle


\section{INTRODUCTION}

Quantum Chromodynamics (QCD) is our theory of the strong interactions.
However, we are far from understanding how it works in many important cases.
Many hadrons discovered recently have puzzling properties.  Hadron spectra
often are crucial in separating electroweak physics from strong-interaction
effects.  QCD may not be the only instance of important non-perturbative
effects; one should be prepared for surprises at the Large Hadron Collider
(LHC).  Sharpening spectroscopic techniques even may help understand the
intricate structure of masses and transitions at the quark and lepton level.

The QCD scale is $\sim 200$ MeV (momentum) or $\sim 1$ fm (distance), where
perturbation theory cannot be used.  Although lattice gauge theories are the
eventual tool of choice for describing effects in this regime, several
other methods can provide information, especially for multi-quark and
multi-hadron problems not yet feasible with lattice techniques.
These include chiral dynamics (treating soft pions, chiral solitons,
and possibly parity doubling in spectra \cite{Jaffe06}), heavy quark symmetry
(describing hadrons with one charm or beauty quark as QCD ``hydrogen'' or
``deuterium'' atoms), studies of correlations among quarks \cite{KL,JW,SW}
and new states they imply (such as a weakly decaying $bq \bar c \bar q'$
state \cite{KL}), potential descriptions (including relativistic and
coupled-channel descriptions), and QCD sum rules.  I will describe phenomena to
which these methods might be applied.

In the present review I treat light-quark (and no-quark) states, charmed
and beauty hadrons, and heavy quarkonium ($c \bar c$ and $b \bar b$), and
conclude with some future prospects.

\section{LIGHT-QUARK STATES}

Several issues are of interest these days in light-quark spectroscopy.  These
include (1) the nature of the low-energy S-wave $\pi \pi$ and $K \pi$
interactions; (2) the proliferation of interesting threshold effects in a
variety of reactions, and (3) the interaction of quark and gluonic degrees
of freedom.
\vskip -0.3in

\subsection{Low-energy $\pi \pi$ S-wave}
\vskip -0.1in

An S-wave $\pi \pi$ low-mass correlation in the $I=0$ channel (``$\sigma$'')
has been used for many years to describe nuclear forces.  Is it a resonance?
What is its quark content?  What can we learn about it from charm and beauty
decays?  This particle, otherwise known as $f_0(600)$ \cite{PDG}, can be
described as a dynamical $I=J=0$ resonance in elastic $\pi \pi$ scattering
using current algebra, crossing symmetry, and unitarity \cite{BG71,VB85,%
Dobado:1992ha}.  It appears as a pole with a large imaginary part with real
part at or below $m_\rho$.  Its effects differ in $\pi \pi \to \pi \pi$, where
an Adler zero suppresses the low-energy amplitude, and inelastic processes
such as $\gamma \gamma \to \pi \pi$ \cite{GR}, where the lack of an Adler
zero leads to larger contributions at low $m_{\pi \pi}$.

\begin{figure}
\includegraphics[width=0.75\textwidth]{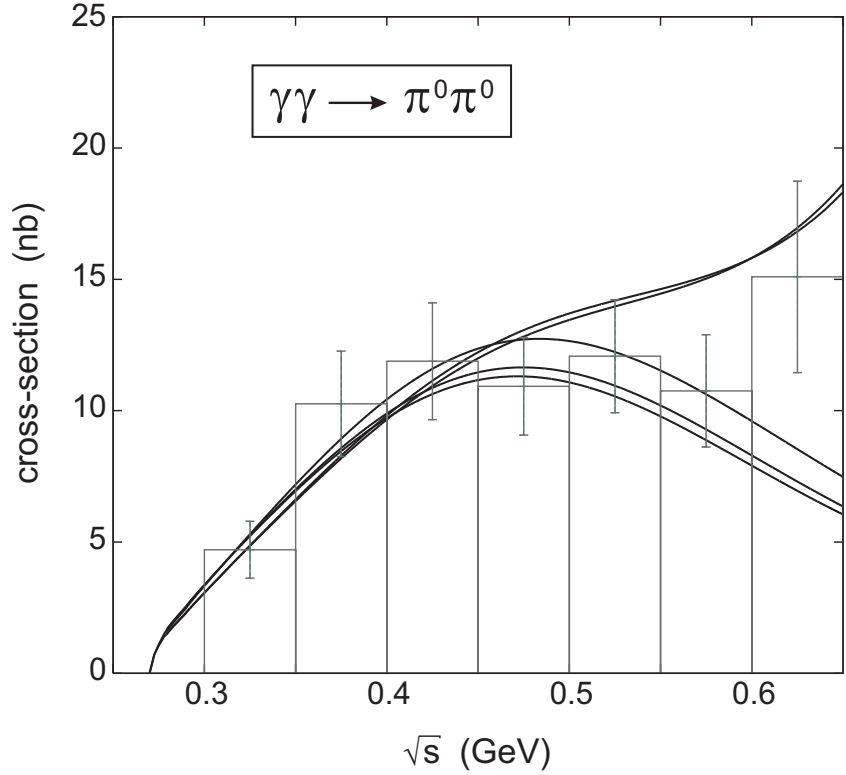}
\caption{Fits to $\gamma \gamma \to \pi^0 \pi^0$ in various models.  From
Ref.\ \cite{MP06}.
\label{fig:ggpzpz}}
\end{figure}

Modern treatments of the low-energy $\pi \pi$ interaction implement crossing
symmetry using an elegant set of exact low-energy relations \cite{Roy71}.  In
one approach \cite{CCL} a $\sigma$ pole is found at $441 - i 272$ MeV,
corresponding to a full width at half maximum of 544 MeV; another
\cite{vanBeveren:2006ua} finds the pole at $555 - i 262$ MeV.  Such a $\sigma$
provides a good description of $\gamma \gamma \to \pi^0 \pi^0$ \cite{MP06},
as shown in Fig.\ \ref{fig:ggpzpz}, with $\Gamma(\sigma \to \gamma \gamma) =
(4.1 \pm 0.3)$ keV.  While this large partial width might be viewed as favoring
a $q \bar q$ interpretation of $\sigma$ \cite{MP06}, a $\pi \pi$ dynamical
resonance seems equally satisfactory \cite{GR}.  Other recent manifestations
of a $\sigma$ include the decays $D^+ \to \sigma \pi^+ \to \pi^+ \pi^- \pi^+$
\cite{E791sigma} and $J/\psi \to \omega \sigma \to \omega \pi^+ \pi^-$
\cite{BESsigma}, where the $\sigma$ pole appears at $(541\pm 39) - i
(252 \pm 42)$ MeV (or $(500 \pm 30) - i(264 \pm 30)$ MeV in an independent
analysis \cite{Bugg:2006gc}).  Successful fits without a $\sigma$ have been
performed, but have been criticized in Ref.\ \cite{Bugg:2005nt}.
\vskip -0.3in

\subsection{Low-energy $K \pi$ S-wave}
\vskip -0.1in

Is there a low-energy $K \pi$ correlation (``$\kappa$'')?  Can it be
generated dynamically in the same manner as the $\sigma$?  Some insights are
provided in \cite{Dobado:1992ha,Oller}.

The low-energy $K \pi$ interaction in the $I=1/2,~J=0$ channel is favorable to
dynamical resonance generation:  The sign of the scattering length is the
same as for the $I=J=0~\pi \pi$ interaction.  A broad scalar resonance $\kappa$
is seen in the $I=1/2,~J=0~K^- \pi^+$ subsystem in $D^+ \to K^-\pi^+  \pi^+$,
and a model-independent phase shift analysis shows resonant $J=0$ behavior in
this subsystem \cite{Aitala:2002kr}.  The $\kappa$ is also seen by the BES II
Collaboration in $J/\psi \to \bar K^{*0}(892) K^+ \pi^-$ decays
\cite{Ablikim:2005ni}.  An independent analysis of the BES II data
\cite{vanBeveren:2006ua} finds a $\kappa$ pole at $745 - i 316$ MeV, while a
combined analysis of $D^+ \to K^-\pi^+  \pi^+$, elastic $K \pi$ scattering,
and the BES II data \cite{Bugg:2005xx} finds a pole at $M(\kappa) =
(750^{+30}_{-55}) - i(342 \pm 60)$ MeV.

\begin{figure}
\includegraphics[height=0.47\textheight]{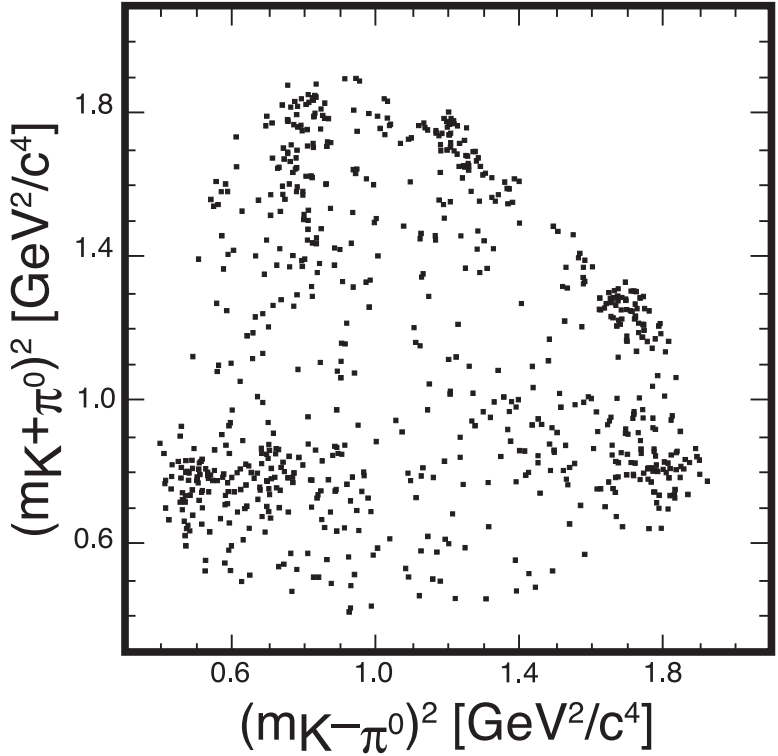}
\caption{Dalitz plot for $D^0 \to K^+ K^- \pi^0$.  From Ref.\ \cite{Paras06}.
\label{fig:KKpi}}
\end{figure}

The $\kappa$, like the $\sigma$, is optional in many descriptions of
final-state interactions.  An example is a recent fit to the $D^0 \to K^+ K^-
\pi^0$ Dalitz plot based on CLEO data \cite{Paras06}, shown in Fig.\
\ref{fig:KKpi}.  The bands correspond to $K^{*-}$ (vertical), $K^{*+}$
(horizontal), and $\phi$ (diagonal).  One can see the effect of an S-wave
(nonresonant or $\kappa$) background interfering with $K^{*+}$ and $K^{*-}$
with opposite signs on the left and bottom of the plot.

Depopulated regions at $m(K^\pm \pi^0) \simeq$ 1 GeV/$c^2$ may be due to the
opening of the $K \pi^0 \to K \eta$ S-wave threshold (a $D^0 \to K^+ K^- \eta$
Dalitz plot would test this) or to a vanishing S-wave $K \pi$ amplitude
between the $\kappa$ and a higher $J^P = 0^+$ resonance.

\subsection{Dips and edges}
\vskip -0.1in

With the advent of high-statistics Dalitz plots for heavy meson decays one
is seeing a number of dips and edges which often are evidence for thresholds
\cite{JLRthr}.
An example is shown in a recent $D^0 \to K_S^0 \pi^+ \pi^-$ plot (Fig.\ 3)
from BaBar \cite{BaKspipi} (see also results from Belle \cite{BeKspipi} and
CLEO \cite{CLKspipi}).  The vertical band corresponds to $K^{*-}$ and the
diagonal to $\rho^0$.  The sharp edges along the diagonal in the $\pi^+ \pi^-$
spectrum correspond to $\rho$-$\omega$ interference [around $M(\pi \pi) = 0.8$
GeV/$c^2$] and to $\pi^+ \pi^- \leftrightarrow K \bar K$ [around $M(\pi \pi) =
1$ GeV/$c^2$].  Rapid variation of an amplitude occurs when a new S-wave
channel opens because no centrifugal barrier is present.

\begin{figure}
\includegraphics[height=0.4\textheight]{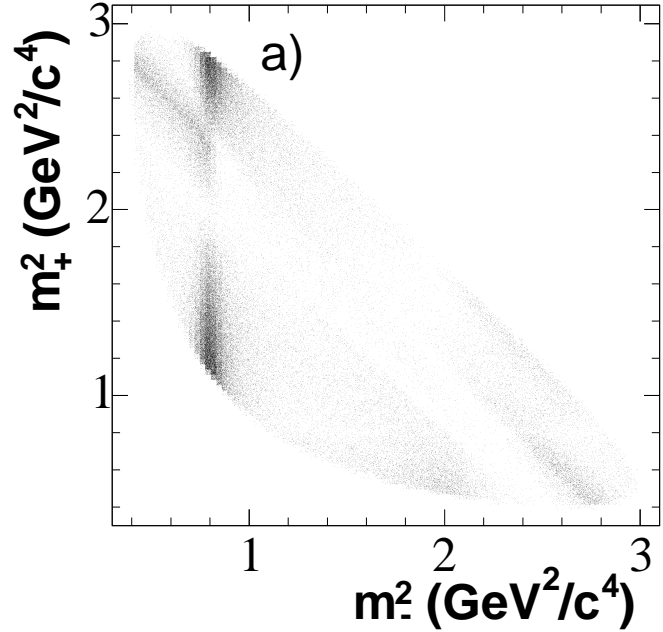}
\caption{Dalitz plot for $D^0 \to K_S^0 \pi^+ \pi^-$.  From Ref.\
\cite{BaKspipi}.
\label{fig:KsPiPi}}
\end{figure}

Further dips are seen in $6 \pi$ photoproduction just at $p \bar p$ threshold;
in $R_{e^+ e^-}$ just below the threshold for S-wave production of $D(1865)
+D_1(2420)$; and in the Dalitz plot for $B^\pm \to K^\pm K^\mp K^\pm$
around $M(K^+ K^-) = 1.6$ GeV/$c^2$ \cite{BaKKK}, which could be a threshold
for vector meson pair production.
\vskip -1in

\subsection{Glueballs and hybrids}
\vskip -0.1in

In QCD, quarkless ``glueballs'' may be constructed from pure-glue
configurations: $F^a_{\mu \nu} F^{a \mu \nu}$ for $J^{PC} = 0^{++}$ states,
$F^a_{\mu \nu} \tilde{F}^{a \mu \nu}$ for $J^{PC} = 0^{-+}$ states, etc., where
$F^a_{\mu \nu}$ is the gluon field-strength tensor.  All such states should be
flavor-singlet with isospin $I=0$, though couplings of spinless states to $s
\bar s$ could be favored \cite{Chanowitz:2005du}.  Lattice QCD calculations
predict the lowest glueball to be $0^{++}$ with $M \simeq 1.7$ GeV
\cite{Campbell:1997}.  The next-lightest states, $2^{++}$ and $0^{-+}$, are
expected to be several hundred MeV/$c^2$ heavier.  Thus it is reassuring
that the lightest mainly flavor-singlet state, the $\eta'$, is only
gluonic $(8 \pm 2)\%$ of the time, as indicated by a recent measurement
of ${\cal B}(\phi \to \eta' \gamma)$ by the KLOE Collaboration \cite{giov}.

Many other $I=0$ levels, e.g., $q \bar q$, $q \bar q g$ ($g$ = gluon), $q q
\bar q \bar q, \ldots$, can mix with glueballs.  One must study $I=0$ levels
and their mesonic couplings to separate out glueball, $n \bar n \equiv (u \bar
u + d \bar d)/ \sqrt{2}$, and $s \bar s$ components.  Understanding the rest of
the {\it flavored} $q \bar q$ spectrum for the same $J^P$ thus is crucial.
The best $0^{++}$ glueball candidates (mixing with $n \bar n$ and $s \bar s$)
are at 1370, 1500, and 1700 MeV.  One can explore their flavor structure
through production and decay, including looking for their $\gamma(\rho,\omega,
\phi)$ decays \cite{ClZh}.  A CLEO search for such states in $\Upsilon(1S)
\to \gamma X$ finds no evidence for them but does see the familiar resonance
$f_2(1270)$ \cite{CLf}.

QCD predicts that in addition to $q \bar q$ states there should be $q \bar q g$
(``hybrid'') states containing a constituent gluon $g$.  One signature of them
would be states with quantum numbers forbidden for $q \bar q$ but allowed for
$q \bar q g$.  For $q \bar q$, $P = (-1)^{L+1},~C = (-1)^{L+S}$, so $CP =
(-1)^{S+1}$.  The forbidden $q \bar q$ states are then those with $J^{PC}=
0^{--}$ and $0^{+-},~1^{-+},~2^{+-},\ldots$.  A consensus in quenched lattice
QCD is that the lightest exotic hybrids have $J^{PC} = 1^{-+}$ and $M(n \bar n
g) \simeq 1.9$ GeV, $M(s \bar s g) \simeq 2.1$ GeV, with errors 0.1--0.2 GeV
\cite{McNeile:2002az}.  (Unquenched QCD must treat mixing with $qq \bar q \bar
q$ and meson pairs.) Candidates for hybrids include $\pi_1(1400)$ (seen in some
$\eta \pi$ final states, e.g., in $p \bar p$ annihilations) and $\pi_1(1600)$
(seen in $3 \pi$, $\rho \pi$, $\eta' \pi$).
Brookhaven experiment E-852 published evidence for a $1^{-+}$ state called
$\pi_1(1600)$ \cite{Adams:1998ff}.  A recent analysis by a subset of E-852's
participants \cite{Dzierba:2005sr} does not require this particle if
a $\pi_2(1670)$ contribution [an orbital excitation of the $\pi(140)$]
is assumed.  The favored decays of a $1^{-+}$ hybrid are to a $q \bar q
(L=0)$ + $q \bar q (L=1)$ pair, such as $\pi b_1(1235)$.  A detailed review of
glueballs and hybrids has been presented by C. Meyer at this Conference
\cite{Meyer}.

\section{CHARMED STATES}

\begin{figure}
\includegraphics[height=0.39\textheight]{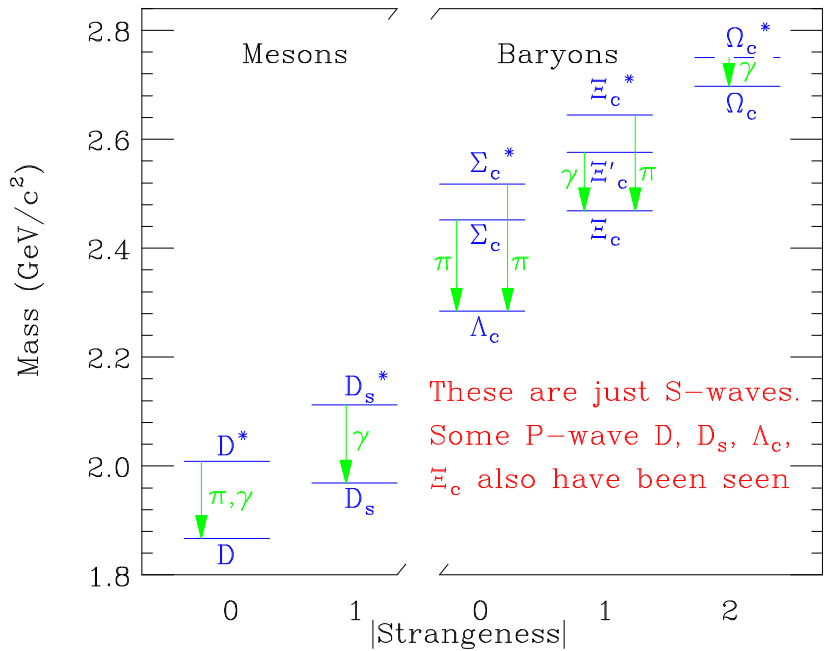}
\caption{Lowest S-wave states with a single charmed quark.  Only the
$\Omega_c^*$ (dashed line denotes predicted mass) has not yet been reported.
\label{fig:charm}}
\end{figure}

The present status of the lowest S-wave states with a single charmed quark is
shown in Fig.\ \ref{fig:charm}.  We will discuss progress on
orbitally-excited charmed baryons \cite{Mallik,Tsuboyama} and charmed-strange
mesons, with brief remarks on $D^+$ and $D_s^+$ decay constants which are
treated in more detail in Ref.\ \cite{Briere}.
\vskip -0.3in

\subsection{Charmed $L > 0$ baryons}
\vskip -0.1in

\begin{figure}
\includegraphics[width=0.46\textheight]{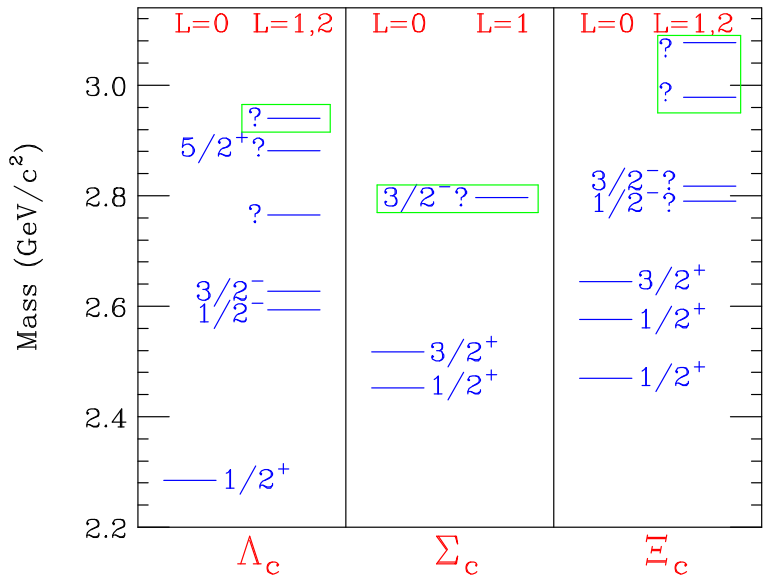}
\caption{Singly-charmed baryons and some of their orbital excitations.
\label{fig:lexb}}
\end{figure}

For many years CLEO was the main source of data on orbitally-excited charmed
baryons.  Now BaBar and Belle are
discovering new states, denoted by the outlined levels in Fig.\ \ref{fig:lexb}.
The Belle Collaboration observed an excited $\Sigma_c$ candidate decaying to
$\Lambda_c \pi^+$, with mass about 510 MeV above $M(\Lambda_c)$
\cite{Mizuk:2004yu}.  The value of its $J^P$ shown in Fig.\ \ref{fig:lexb} is a
guess, using the diquark ideas of \cite{SW}.  The highest $\Xi_c$ levels were
reported by Belle in Ref.\ \cite{Becb}.  The highest $\Lambda_c$ is seen
by BaBar in the decay mode $D^0 p$ \cite{Bacb}.

In Fig.\ \ref{fig:lexb} the first excitations of the $\Lambda_c$ and $\Xi_c$
are similar, scaling well from the first $\Lambda$ excitations
$\Lambda(1405,1/2^-)$ and $\Lambda(1520,3/2^-)$.  They have the same cost in
$\Delta L$ (about 300 MeV), and their $L \cdot S$ splittings scale as $1/m_s$
or $1/m_c$.  Higher $\Lambda_c$ states may correspond to excitation of a
spin-zero $[ud]$ pair to $S=L=1$, leading to many allowed $J^P$ values up to
$5/2^-$.  In $\Sigma_c$ the light-quark pair has $S=1$; adding $L=1$ allows
$J^P \le 5/2^-$.  States with higher $L$ may be narrower as a result of
inreased barrier factors affecting their decays, but genuine spin-parity
analyses would be very valuable.

\subsection{Lowest charmed-strange $0^+$, $1^+$ states}
\vskip -0.1in

\begin{figure}
\includegraphics[width=0.95\textwidth]{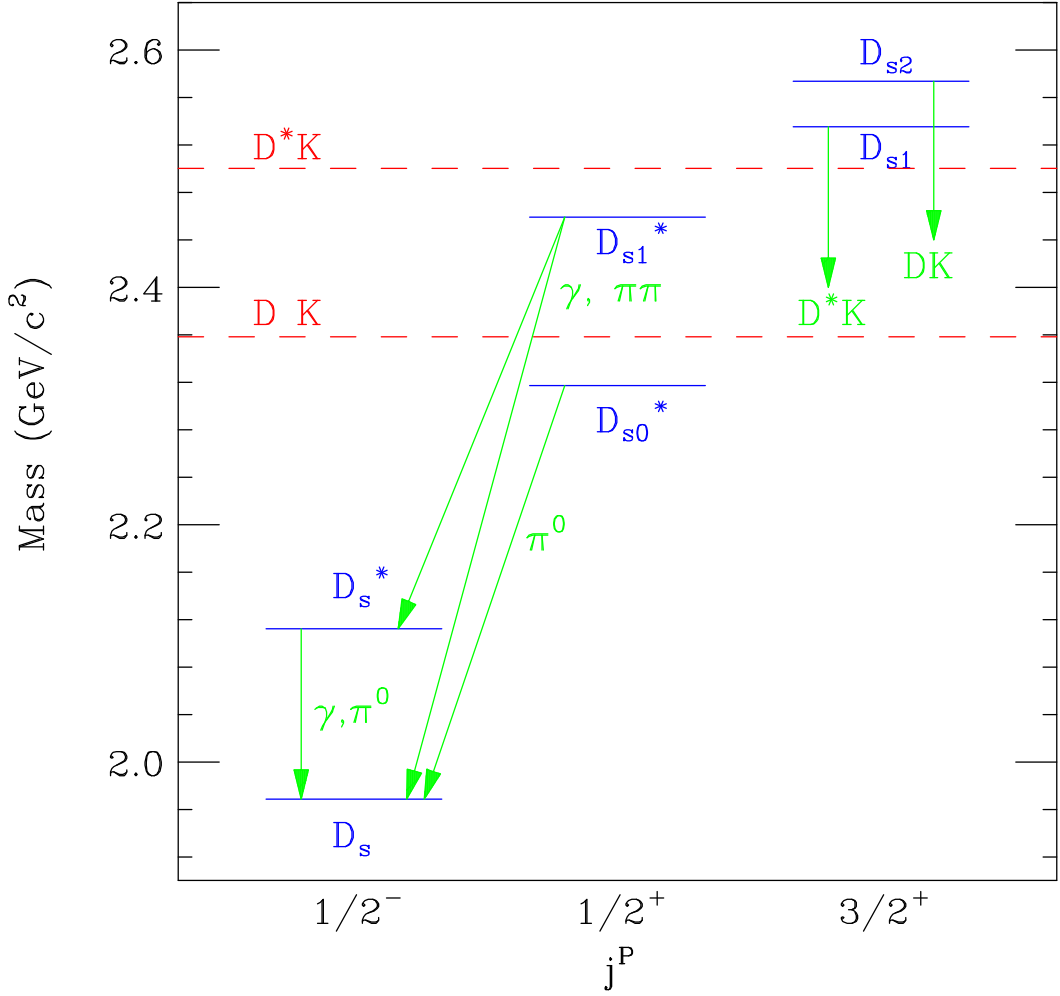}
\caption{Charmed-strange mesons with $L=0$ (negative-parity) and $L=1$
(positive-parity).  Here $j^P$ denotes the total light-quark
spin + orbital angular momentum and the parity $P$.
\label{fig:ds}}
\end{figure}

In the past couple of years the lowest $J^P = 0^+$ and $1^+$ $c \bar s$
states turned out to have masses well below most expectations.  If they had
been as heavy as the already-seen $c \bar s$ states with $L=1$, the
$D_{s1}(2536)$ [$J^P = 1^+$] and $D_{s2}(2573)$ [$J^P = 2^+$]), they would
have been able to decay to $D \bar K$ (the $0^+$ state) and $D^* \bar K$ (the
$1^+$ state).  Instead several groups \cite{Aubert:2003fg} observed a narrow
$D_s(2317) \equiv D_{s0}^*$ decaying to $\pi^0 D_s$ and a narrow $D_s(2460)
\equiv D_{s1}^*$ decaying to $\pi^0 D_s^*$, as illustrated in Fig.\
\ref{fig:ds}.  Their low masses allow the isospin-violating and electromagnetic
decays of $D_{s0}^*$ and $D_{s1}^*$ to be observable.  The decays $D_s(2460)
\to D_s \gamma$ and $D_s(2460) \to D_s \pi^+ \pi^-$ also have been seen
\cite{Mallik,Marsiske}, and the absolute branching ratios
${\cal B}(D_{s1}^* \to \pi^0 D_s^*) = (0.56 \pm 0.13 \pm 0.09)\%,$
${\cal B}(D_{s1}^* \to \gamma D_s) = (0.16 \pm 0.04 \pm 0.03)\%,$
${\cal B}(D_{s1}^* \to \pi^+ \pi^- D_s^*) = (0.04 \pm 0.01)\%$
measured.

The selection rules in decays of these states show their $J^P$ values
are consistent with $0^+$ and $1^+$.  Low masses are predicted
\cite{Bardeen:2003kt} if these states are viewed as parity-doublets of the
$D_s(0^-)$ and $D^*_s(1^-)$ $c \bar s$ ground states in the framework of
chiral symmetry.  The splitting from the ground states is 350 MeV in each case.
Alternatively, one can view these particles as bound states of $D^{(*)}K$,
perhaps bound by the transitions $(c \bar q)(q \bar s) \leftrightarrow (c \bar
s)$ (the binding energy in each case would be 41 MeV), or as $c \bar s$ states
with masses lowered by coupling to $D^{(*)}K$ channels
\cite{vanBeveren:2003kd,Close:2004ip,NewDs}.

\subsection{$D^+$ and $D_s$ decay constants}
\vskip -0.1in

CLEO has reported the first significant measurement of the $D^+$ decay
constant: $f_{D^+} = (222.6 \pm 16.7^{+2.8}_{-3.4})$ MeV \cite{Artuso:2005ym}.
This is consistent with lattice predictions, including one \cite{Aubin:2005ar}
of $(201 \pm 3 \pm 17)$ MeV.  The accuracy of the previous world average
\cite{PDG} $f_{D_s} = (267 \pm 33)$ MeV has been improved by a BaBar value
$f_{D_s} = 283 \pm 17 \pm 7 \pm 14$ MeV \cite{Aubert:2006sd} and a new
CLEO value $f_{D_s} = 280.1 \pm 11.6 \pm 6.0$ MeV
\cite{Stone06}.  The latter,
when combined with CLEO's $f_D$, leads to $f_{D_s}/f_D = 1.26 \pm 0.11 \pm
0.03$.  A lattice prediction for $f_{D_s}$ \cite{Aubin:2005ar} is $f_{D_s} =
249 \pm 3 \pm 16$ MeV, leading to $f_{D_s}/f_D = 1.24 \pm 0.01 \pm 0.07$.
One expects $f_{B_s}/f_B \simeq f_{D_s}/f_D$ so better measurements
of $f_{D_s}$ and $f_D$ by CLEO will help validate lattice calculations and
provide input for interpreting $B_s$ mixing.  A desirable error on $f_{B_s}/f_B
\simeq f_{D_s}/f_D$ is $\le 5\%$ for useful determination of CKM element ratio
$|V_{td}/V_{ts}|$, needing errors $\le 10$ MeV on $f_{D_s}$ and $f_D$.
The ratio $|V_{td}/V_{ts}| = 0.208^{+0.008}_{-0.006}$ is implied by a recent
CDF result on $B_s$--$\overline{B}_s$ mixing \cite{CDFmix} combined with
$B$--$\overline{B}$ mixing and $\xi \equiv (f_{B_s} \sqrt{B_{B_s}}/f_B
\sqrt{B_B}) = 1.21^{+0.047}_{-0.035}$ from the lattice \cite{Okamoto}.
A simple quark model scaling argument anticipated $f_{D_s}/f_D \simeq
f_{B_s}/f_B \simeq \sqrt{m_s/m_d} \simeq 1.25$, where $m_s \simeq 485$ MeV and
$m_d \simeq 310$ MeV are constituent quark masses \cite{Rosner90}.

\section{BEAUTY HADRONS}

\begin{figure}
\includegraphics[width=0.98\textwidth]{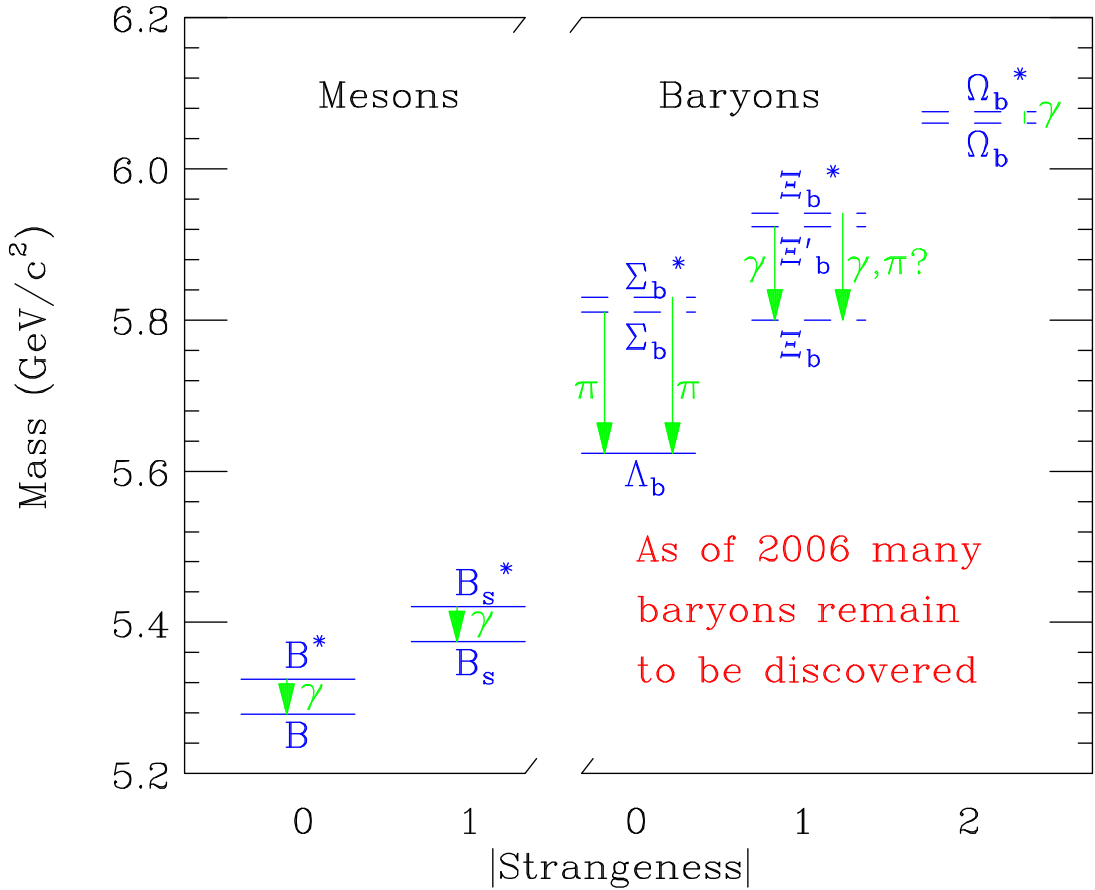}
\caption{S-wave hadrons containing a single beauty quark.  Dashed lines denote
predicted levels not yet observed.
\label{fig:beauty}}
\end{figure}

The spectrum of ground-state hadrons containing a single $b$ quark is shown in
Fig.\ \ref{fig:beauty}.  The following are a few recent high points of beauty
hadron spectroscopy.

The CDF Collaboration has identified events of the form $B_c \to J/\psi
\pi^\pm$, allowing for the first time a precise determination of the
mass: $M$=(6276.5$\pm$4.0$\pm$2.7) MeV/$c^2$ \cite{Aoki:2006}.  This is in
reasonable accord with the latest lattice prediction of
6304$\pm$12$^{+18}_{-0}$ MeV \cite{Allison:2004be}.

The long-awaited $B_s$--$\overline{B}_s$ mixing has finally been observed
\cite{CDFmix,D0mix}.  The CDF value, $\Delta m_s = 17.31^{+0.33}_{-0.18} \pm
0.07$ ps$^{-1}$, constrains $f_{B_s}$ and $|V_{td}/V_{ts}|$, as mentioned
earlier.

The Belle Collaboration has observed the decay $B \to \tau \nu_\tau$
\cite{Btaunu}, leading to $f_B |V_{ub}| = (7.73^{+1.24+0.66}_{-1.02-0.58})
\times 10^{-4}$ GeV.  When combined with an estimate \cite{HL06} $f_{B_d} =
(191 \pm 27)$ MeV, this leads to $|V_{ub}| = (4.05 \pm 0.89) \times 10^{-3}$,
which is squarely in the range of recent averages \cite{HFAG}.

A new CDF value for the $\Lambda_b$ lifetime, $\tau(\Lambda_b) = (1.59
\pm 0.08 \pm 0.03)$ ps, was reported at this Conference \cite{CDFLam}.  Whereas
the previous world average of $\tau(\Lambda_b)$ was about 0.8 that of $B^0$,
below theoretical predictions, the new CDF value substantially increases the
world average to a value $\tau(\Lambda_b) = (1.410 \pm 0.054)$ ps which is
$0.923 \pm 0.036$ that of $B^0$ and quite comfortable with theory.

\section{CHARMONIUM}
\vskip -0.1in

\subsection{Observation of the $h_c$}
\vskip -0.1in

\begin{figure}
\includegraphics[width=0.9\textwidth]{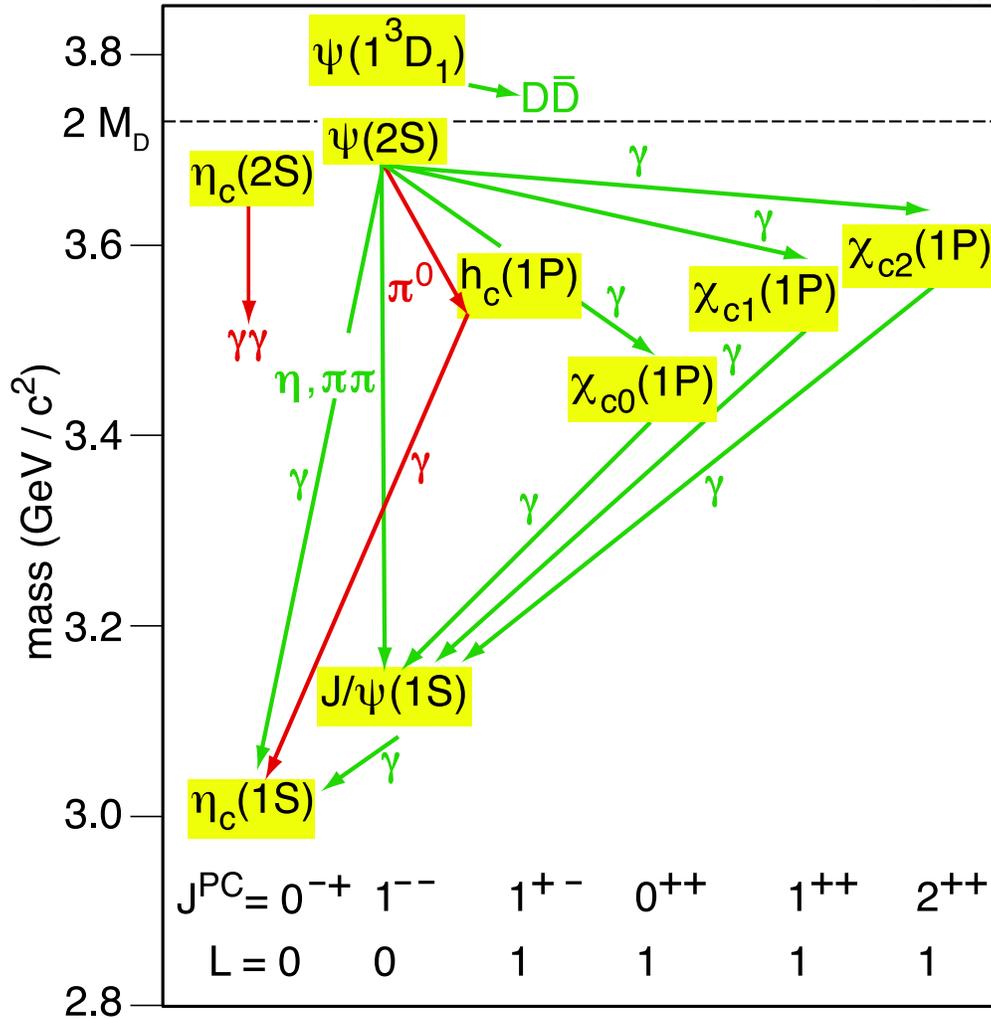}
\caption{Transitions among low-lying charmonium states.  From Ref.\
\cite{ccour}.
\label{fig:ccour}}
\end{figure}

The $h_c(1^1P_1)$ state of charmonium has been observed by CLEO
\cite{Rosner:2005ry,Rubin:2005px} via $\psi(2S) \to \pi^0 h_c$ with $h_c
\to \gamma \eta_c$ (transitions denoted by red (dark) arrows in
Fig.\ \ref{fig:ccour} \cite{ccour}).

Hyperfine splittings test the spin-dependence and spatial behavior of the $Q
\bar Q$ force.  Whereas these splittings are $M(J/\psi) - M(\eta_c) \simeq 115$
MeV for 1S and $M[\psi'] - M(\eta'_c) \simeq $49 MeV for 2S levels, P-wave
splittings should be less than a few MeV since the potential is proportional to
$\delta^3(\vec{r})$ for a Coulomb-like $c \bar c$ interaction.  Lattice QCD
\cite{latt} and relativistic potential \cite{Ebert:2002pp} calculations confirm
this expectation.  One expects $M(h_c) \equiv M(1^1P_1) \simeq
\langle M(^3P_J) \rangle = 3525.36 \pm 0.06$ MeV.

Earlier $h_c$ sightings \cite{Rosner:2005ry,Rubin:2005px} based on
$\bar p p$ production in the direct channel, include a few events at $3525.4
\pm 0.8$ MeV seen in CERN ISR Experiment R704; a state at $3526.2 \pm
0.15 \pm 0.2$ MeV, decaying to $\pi^0 J/\psi$, reported by Fermilab E760 but
not confirmed by Fermilab E835; and a state at $3525.8 \pm 0.2 \pm 0.2$ MeV,
decaying to $\gamma \eta_c$ with $\eta_c \to \gamma \gamma$, reported by
E835 with about a dozen candidate events \cite{Andreotti:2005vu}.

\begin{figure}
\mbox{
\includegraphics[width=0.59\textwidth]{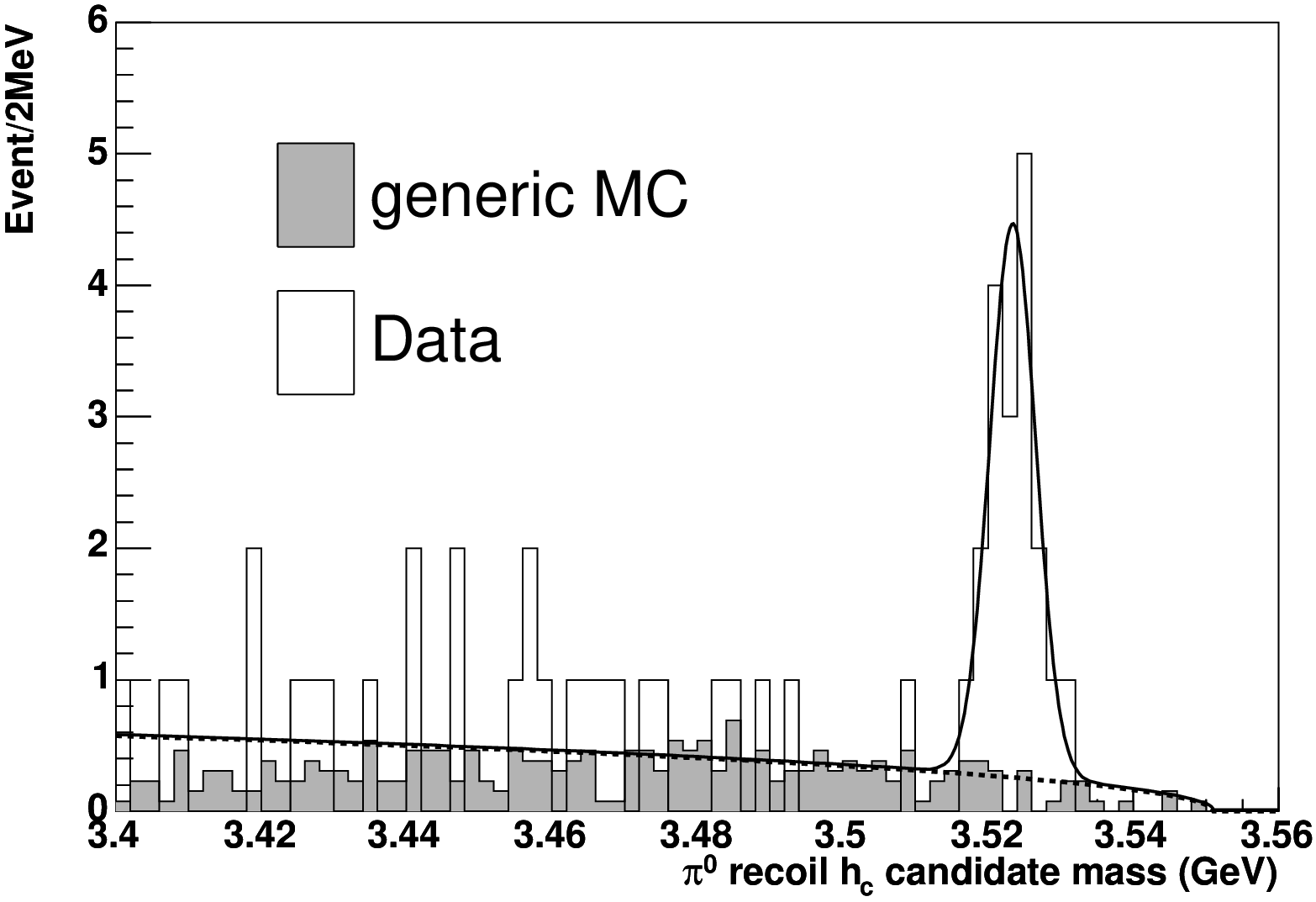}
\includegraphics[width=0.41\textwidth]{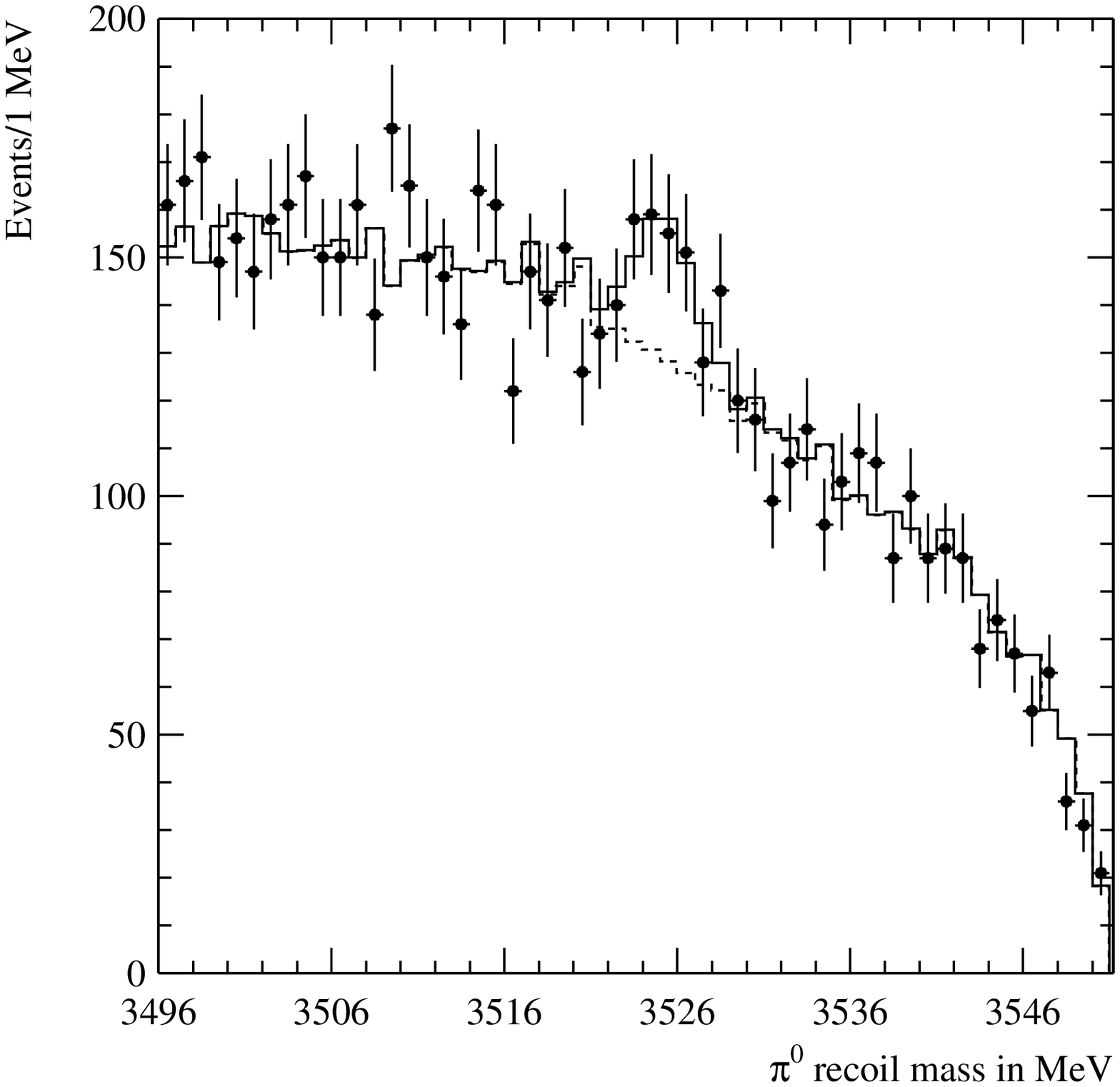}
}
\caption{Left: Exclusive $h_c$ signal from CLEO (3 million $\psi(2S)$
decays).  Data events correspond to open histogram;
Monte Carlo background estimate is denoted by shaded histogram.
The signal shape is a double Gaussian, obtained from signal Monte Carlo.
The background shape is an ARGUS function.
Right: Inclusive $h_c$ signal from CLEO (3 million $\psi(2S)$ decays).
The curve denotes the background function based on generic Monte Carlo plus
signal.  The dashed line shows the contribution of background alone.
Both figures are from Ref.\ \cite{Rubin:2005px}.
\label{fig:hc}}
\end{figure}

In the CLEO data, both inclusive and exclusive analyses see a signal near
$\langle M(^3P_J) \rangle$.  The exclusive analysis reconstructs $\eta_c$ in 7
decay modes, while no $\eta_c$ reconstruction is performed in the inclusive
analysis.
The exclusive signal is shown on the left in Fig.\ \ref{fig:hc}.  A total of 19
candidates were identified, with a signal of $17.5 \pm 4.5$ events above
background.  The mass and product branching ratio for the two transitions 
are $M(h_c) = (3523.6 \pm 0.9 \pm 0.5)$ MeV; ${\cal B}_1(\psi' \to \pi^0 h_c)
{\cal B}_2(h_c \to \gamma \eta_c) = (5.3 \pm 1.5 \pm 1.0) \times 10^{-4}$.
The result of one of two inclusive analyses is shown on the right in
Fig.\ \ref{fig:hc}.  These yield $M(h_c) = (3524.9 \pm 0.7 \pm 0.4)$ MeV,
${\cal B}_1 {\cal B}_2 = (3.5 \pm 1.0 \pm 0.7) \times 10^{-4}$.  Combining
exclusive and inclusive results yields $M(h_c) = (3524.4 \pm 0.6 \pm 0.4)$ MeV,
${\cal B}_1 {\cal B}_2 = (4.0 \pm 0.8 \pm 0.7) \times 10^{-4}$.  The $h_c$ mass
is $(1.0 \pm 0.6 \pm 0.4)$ MeV below $\langle M(^3P_J) \rangle$, barely
consistent with the (nonrelativistic) bound \cite{Stubbe:1991qw} $M(h_c) \ge
\langle M(^3P_J) \rangle$ and indicating little P-wave hyperfine splitting in
charmonium.  The value of ${\cal B}_1 {\cal B}_2$ agrees with theoretical
estimates of $(10^{-3} \cdot 0.4)$.
\vskip -0.3in

\subsection{Decays of the $\psi'' \equiv \psi(3770)$}

The $\psi''(3770)$ is a potential ``charm factory'' for present and future $e^+
e^-$ experiments.  At one time $\sigma(e^+ e^- \to \psi'')$ seemed
larger than $\sigma(e^+ e^- \to \psi'' \to D \bar D)$, raising the question
of whether there were significant non-$D \bar D$ decays of the $\psi''$
\cite{Rosner:2004wy}.  A new CLEO measurement \cite{CLDDbar}, $\sigma(\psi'') =
(6.38 \pm 0.08 ^{+0.41}_{-0.30})$ nb, appears very close to the CLEO value
$\sigma(D \bar D) = 6.39\pm0.10^{+0.17}_{-0.08})$ nb \cite{Briere}, leaving
little room for non-$D \bar D$ decays.  Some question has nonetheless been
raised by two very new BES analyses \cite{BESsig} in which a significant non-$D
\bar D$ component could still be present.

One finds that ${\cal B}(\psi''\to \pi \pi J/\psi,~\gamma \chi_{cJ}, \ldots)$
sum to at most 1--2\%.  Moreover, both CLEO and BES \cite{LP123}, in searching
for enhanced light-hadron modes, find only that the $\rho \pi$ mode,
suppressed in $\psi(2S)$ decays, also is {\it suppressed} in $\psi''$ decays.

Some branching ratios for $\psi'' \to X J/\psi$ \cite{Adam:2005mr} are
${\cal B}(\psi'' \to \pi^+ \pi^- J/\psi) =(0.189\pm0.020\pm0.020)\%$,
${\cal B}(\psi'' \to \pi^0 \pi^0 J/\psi) =(0.080\pm0.025\pm0.016)\%$,
${\cal B}(\psi'' \to \eta J/\psi) = (0.087\pm0.033\pm0.022)\%$, and
${\cal B}(\psi'' \to \pi^0 J/\psi) < 0.028\%$.
The value of ${\cal B}[\psi''(3770) \to \pi^+ \pi^- J/\psi]$ found by CLEO is a
bit above 1/2 that reported by BES \cite{Bai:2003hv}.
These account for less than 1/2\% of the total $\psi''$ decays.

\begin{table}
\caption{CLEO results on radiative decays $\psi'' \to \gamma \chi_{cJ}$.
Theoretical predictions of \cite{Eichten:2004uh} are (a) without and
(b) with coupled-channel effects; (c) shows predictions of
\cite{Rosner:2004wy}.
\label{tab:psipprad}}
\begin{tabular}{ccccc} \hline
Mode & \multicolumn{3}{c}{Predicted (keV)} & CLEO \\
     & (a) & (b) & (c) & \cite{Briere:2006ff} \\ \hline
$\gamma \chi_{c2}$ & 3.2 & 3.9 & 24$\pm$4 & $<21$ \\
$\gamma \chi_{c1}$ & 183 & 59 & $73\pm9$ & $75\pm18$ \\
$\gamma \chi_{c0}$ & 254 & 225 & 523$\pm$12 & $172\pm30$ \\ \hline
\end{tabular}
\end{table}

CLEO has recently reported results on $\psi'' \to \gamma \chi_{cJ}$ partial
widths, based on the exclusive process $\psi'' \to \gamma \chi_{c1,2} \to
\gamma \gamma J/\psi \to \gamma \gamma \ell^+ \ell^-$ \cite{Coan:2005} and
reconstruction of exclusive $\chi_{cJ}$ decays \cite{Briere:2006ff}.  The
results are shown in Table \ref{tab:psipprad}, implying
$\sum_J{\cal B}(\psi'' \to \gamma \chi_{cJ}) = {\cal O}$(1\%).

Several searches for $\psi''(3770) \to ({\rm light~ hadrons})$, including VP,
$K_L K_S$, and multi-body final states have been performed.  Two CLEO analyses
\cite{Adams:2005ks,Huang:2005} find no evidence for any light-hadron $\psi''$
mode above expectations from continuum production except $\phi \eta$,
indicating no obvious signature of non-$D \bar D$ $\psi''$ decays.
\vskip -0.3in

\subsection{$X(3872)$: A $1^{++}$ molecule}
\vskip -0.1in

\begin{figure}
\includegraphics[width=0.99\textwidth]{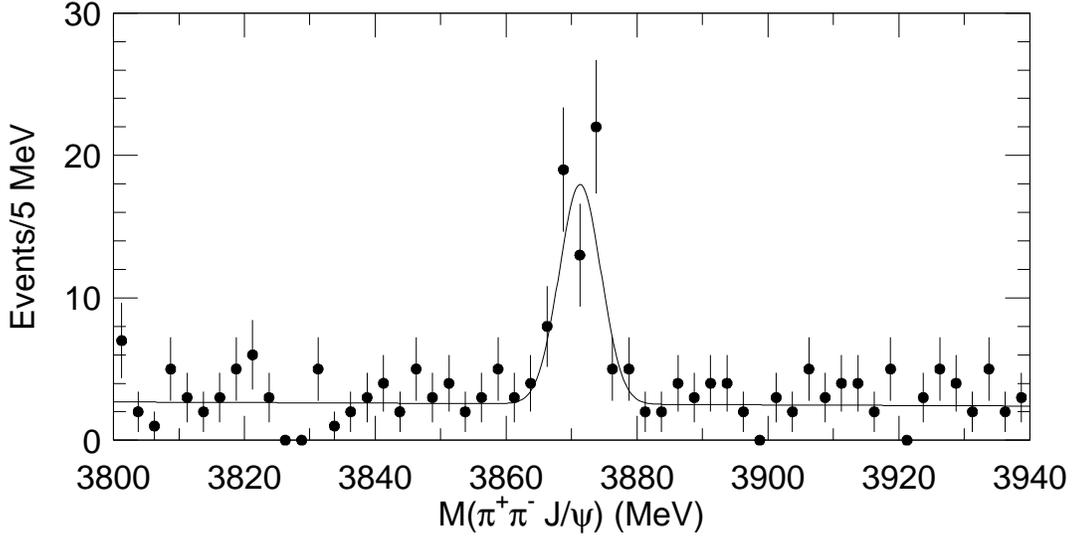}
\caption{Belle distribution in $M(\pi^+ \pi^- J/\psi)$ for the $X(3872)$
region \cite{Abe:2005iy}.
\label{fig:X3872}}
\end{figure}

Many charmonium states above $D \bar D$ threshold have been seen recently.
Reviews may be found in Refs.\ \cite{GodfreyFPCP,Swanson}.
The $X(3872)$, discovered by Belle in $B$ decays \cite{Choi:2003ue}
and confirmed by BaBar \cite{Aubert:2004ns} and in hadronic production
\cite{Acosta:2003zx}, decays predominantly into $J/\psi \pi^+
\pi^-$.  Evidence for it is shown in Fig.\ \ref{fig:X3872} \cite{Abe:2005iy}.
Since it lies well
above $D \bar D$ threshold but is narrower than experimental resolution (a few
MeV), unnatural $J^P = 0^-,1^+, 2^-$ is favored.  It has many features in
common with an S-wave bound state of $(D^0 \bar D^{*0} + \bar D^0 D^{*0})/
\sqrt{2} \sim c \bar c u \bar u$ with $J^{PC} = 1^{++}$ \cite{Close:2003sg}.
The simultaneous decay of $X(3872)$ to $\rho J/\psi$ and $\omega J/\psi$ with
roughly equal branching ratios is a consequence of this ``molecular''
assignment.

Analysis of angular distributions \cite{Rosner:2004ac} in $X \to \rho J/\psi,
\omega J/\psi$ favors the $1^{++}$ assignment \cite{Abe:2005iy}.  (See also
\cite{Marsiske,Swanson}.) The detection of the $\gamma J/\psi$ mode ($\sim
14\%$ of $J/\psi \pi^+ \pi^-$) \cite{Abe:2005ix} confirms the assignment of
positive $C$ and suggests a $c \bar c$ admixture in the wave function.  BaBar
\cite{Bapipipsi} finds ${\cal B}[X(3872) \to \pi^+ \pi^- J/\psi] > 0.042$ at
90\% c.l.

\subsection{Additional states around 3940 MeV}
\vskip -0.1in

\begin{figure}
\mbox{\includegraphics[width=0.39\textwidth]{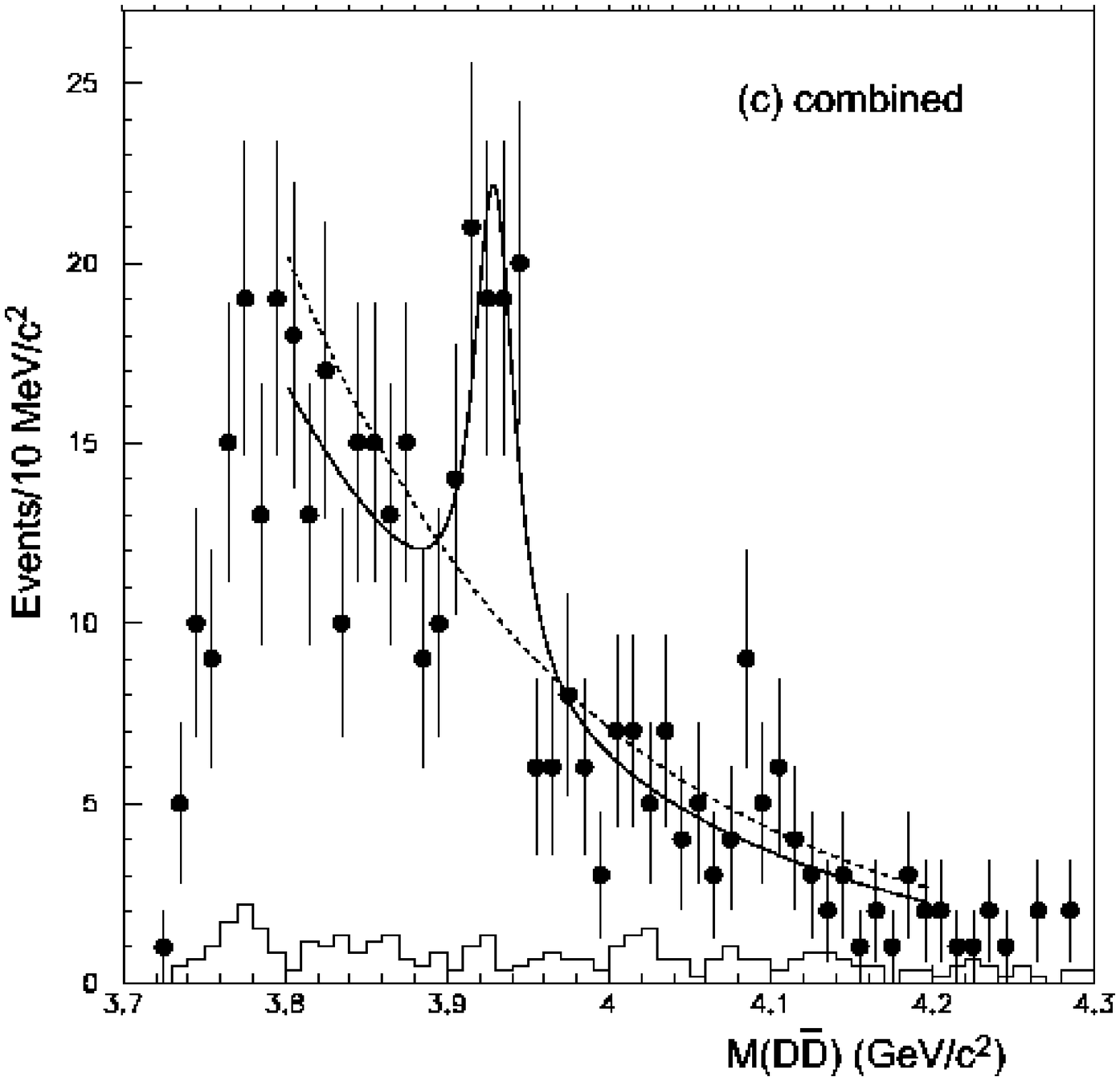}
\includegraphics[width=0.60\textwidth]{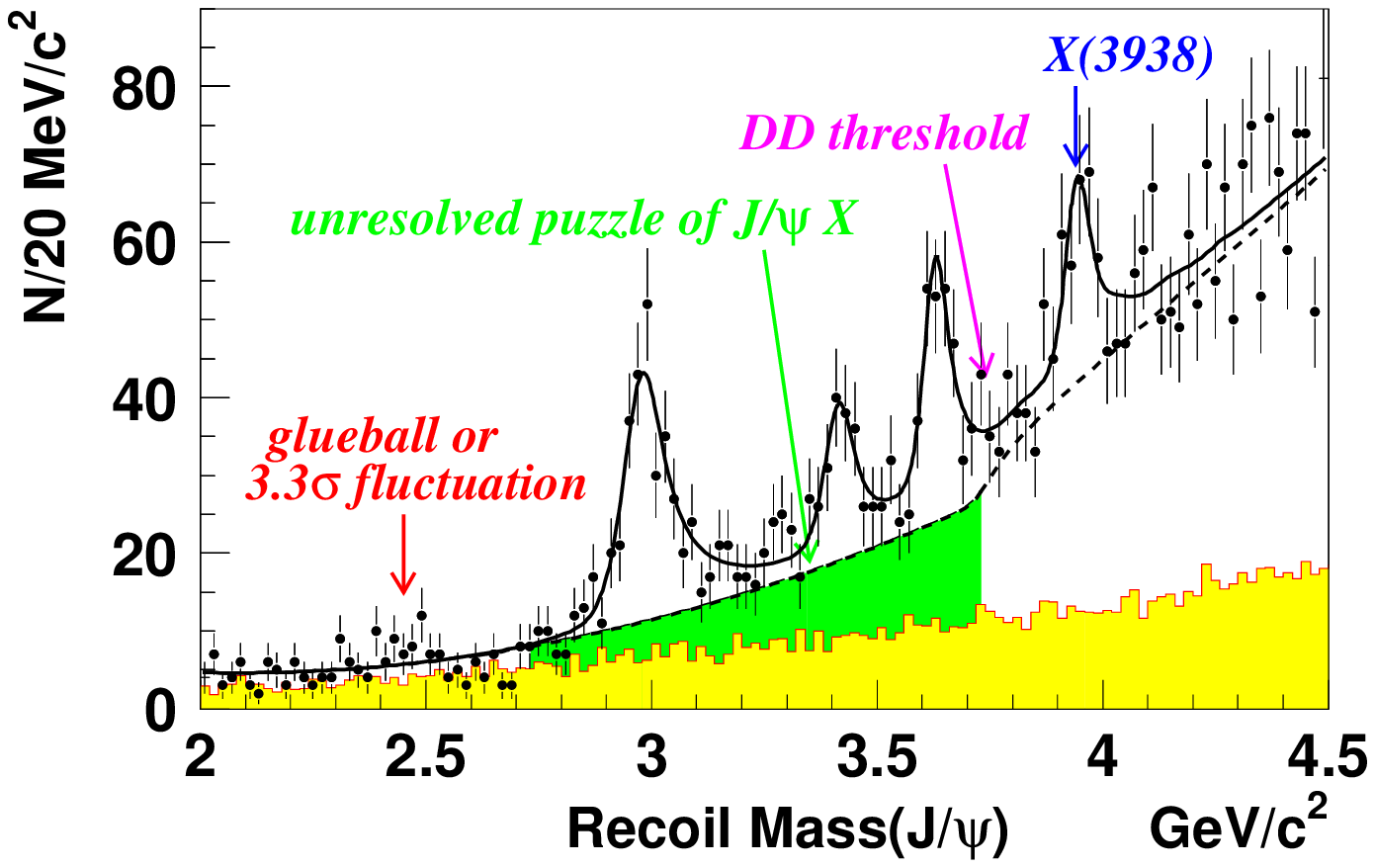}
}
\caption{Left: evidence for an excited $2^3P_2(\chi'_{c2})$ state
(combined $D^0 \bar D^0$ and $D^+ D^-$ spectrum) \cite{Abe:2005bp}.
Right:  Spectrum of masses recoiling against $J/\psi$ in $e^+ e^- \to
J/\psi + X$ \cite{Pakhlov:2004au}.
\label{fig:3940}}
\end{figure}

Belle has reported a candidate for a $2^3P_2(\chi'_{c2})$ state in $\gamma
\gamma$ collisions \cite{Abe:2005bp}, decaying to $D \bar D$ (left panel of
Fig.\ \ref{fig:3940}).  The angular distribution of $D \bar D$ pairs is
consistent with $\sin^4 \theta^*$ as expected for a state with $J=2, \lambda =
\pm2$.  It has $M = 3929 \pm 5 \pm 2$ MeV, $\Gamma = 29 \pm 10 \pm 3$ MeV, and
$\Gamma_{ee} {\cal B}(D \bar D) = 0.18 \pm 0.06 \pm 0.03$ eV, all reasonable
for a $\chi'_{c2}$ state.


A charmonium state $X(3938)$ (the right-most peak in the right panel of Fig.\
\ref{fig:3940}) is produced recoiling against $J/\psi$ in $e^+ e^- \to J/\psi +
X$ \cite{Pakhlov:2004au} and is seen decaying to $D \bar D^*$ + c.c.  Since
all lower-mass states
observed in this recoil process have $J=0$ (these are the $\eta_c(1S),
\chi_{c0}$ and $\eta'_c(2S)$; see the Figure), it is tempting to identify this
state with $\eta_c(3S)$ (not $\chi'_{c0}$, which would decay to $D \bar D$).

The $\omega J/\psi$ final state in $B \to K \omega J/\psi$ shows a peak above
threshold at $M(\omega J/\psi) \simeq 3940$ MeV \cite{Abe:2004zs}.  This could
be a candidate for one or more excited P-wave charmonium states, likely the
$\chi'_{c1,2}(2^3P_{1,2})$.  The corresponding $b \bar b$ states $\chi'_{b1,2}$
have been seen to decay to $\omega \Upsilon(1S)$ \cite{Severini:2003qw}.
\vskip -0.3in

\subsection{The $Y(4260)$}
\vskip -0.1in

\begin{figure}
\includegraphics[width=0.96\textwidth]{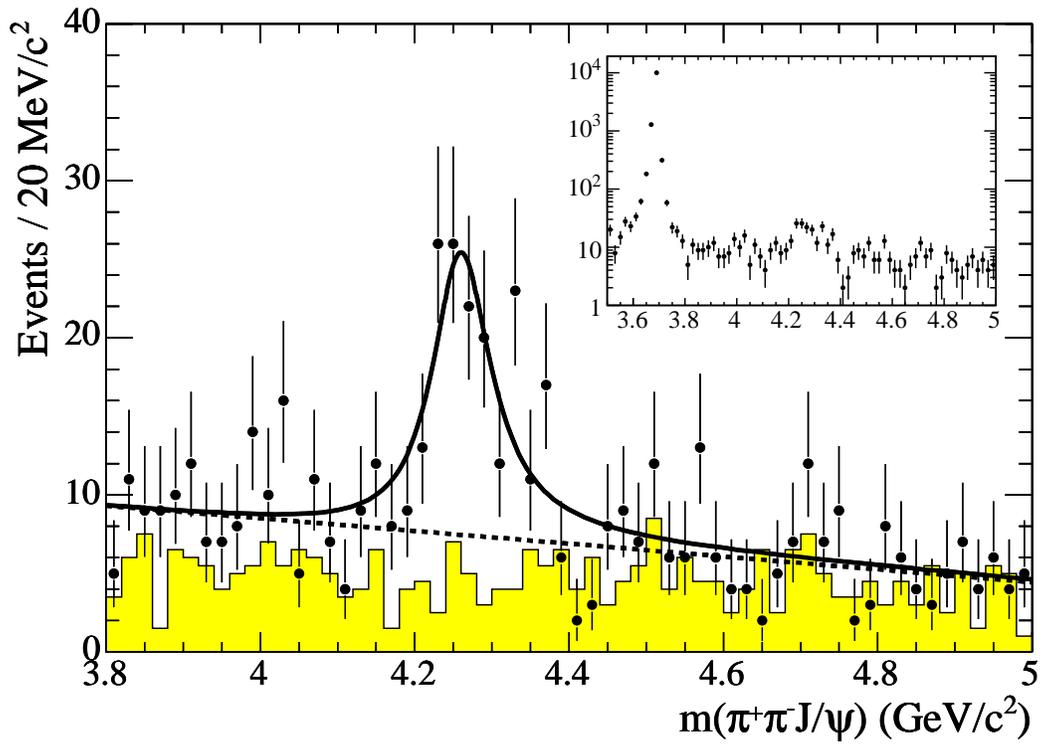}
\caption{Evidence for the $Y(4260)$ \cite{Aubert:2005rm}.
\label{fig:Ba4260}}
\end{figure}

\begin{figure}
\includegraphics[height=0.44\textheight]{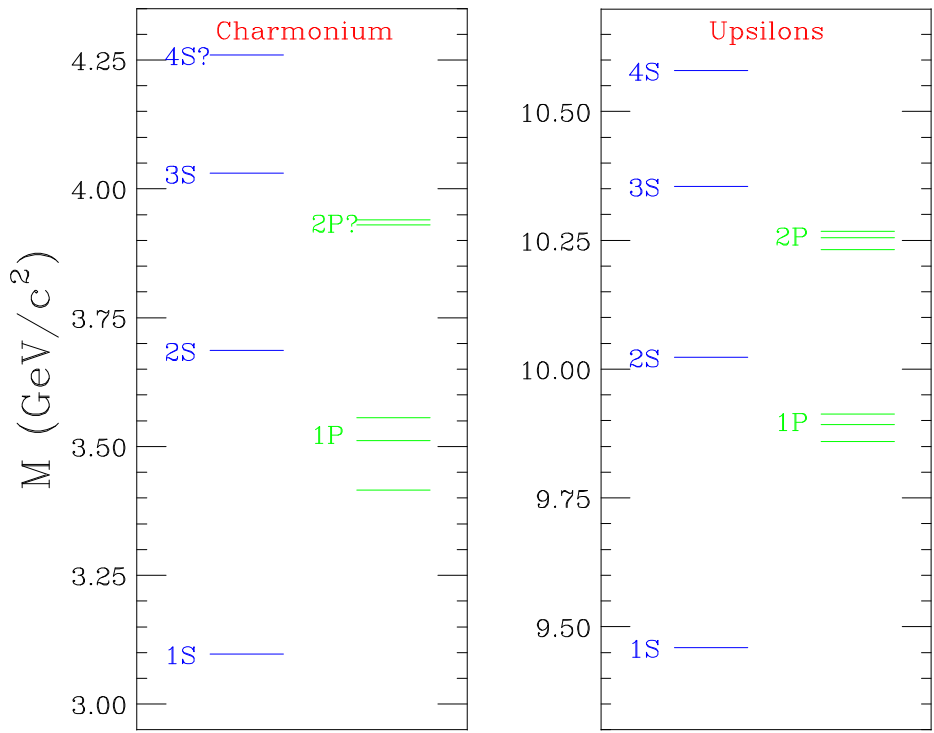}
\caption{Congruence of charmonium and bottomonium spectra if the $Y(4260)$
is a 4S level.
\label{fig:comp}}
\end{figure}

Last year BaBar reported a state $Y(4260)$ produced in the radiative return
reaction $e^+ e^- \to \gamma \pi^+ \pi^- J/\psi$ and seen in the $\pi^+ \pi^-
J/\psi$ spectrum \cite{Aubert:2005rm} (see Fig.\ \ref{fig:Ba4260}).  Its mass
is consistent with being a $4S$ level \cite{Llanes-Estrada:2005vf} since it
lies about 230 MeV above the $3S$ candidate (to be compared with a similar
$4S$-$3S$ spacing in the $\Upsilon$ system).  Indeed, a $4S$ charmonium level
at 4260 MeV/$c^2$ was anticipated on exactly this basis \cite{Quigg:1977dd}.
With this assignment, the $nS$ levels of charmonium and bottomonium are
remarkably congruent to one another, as shown in Fig.\ \ref{fig:comp}.  Their
spacings would be identical if the interquark potential were $V(r) \sim {\rm
log}(r)$, which may be viewed as an interpolation between the short-distance
$\sim -1/r$ and long-distance $\sim r$ behavior expected in QCD.  Other
interpretations of $Y(4260)$ include a $c s \bar c \bar s$ state
\cite{Maiani:2005pe} and a hybrid $c \bar c g$ state \cite{Zhu:2005hp},
for which it lies in the expected mass range.

\begin{figure}
\includegraphics[width=0.4\textheight]{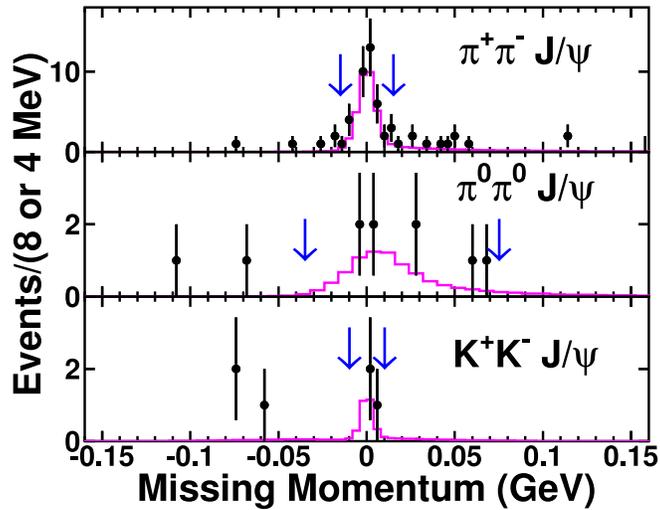}
\caption{Evidence for $Y(4260)$ from a direct scan by CLEO \cite{Coan:2006rv}.
\label{fig:cleo4260}}
\end{figure}

The CLEO Collaboration has confirmed the $Y(4260)$, both in a direct scan
\cite{Coan:2006rv} and in radiative return \cite{Blusk}.  Results from the
scan are shown in Fig.\ \ref{fig:cleo4260}, including signals for $Y(4260)
\to \pi^+ \pi^- J/\psi$ 11$\sigma$), $\pi^0 \pi^0 J/\psi$ (5.1$\sigma$), and
$K^+ K^- J/\psi$ (3.7$\sigma$).  There are also weak signals for $\psi(4160)
\to \pi^+ \pi^- J/\psi$ (3.6$\sigma$) and $\pi^0 \pi^0 J/\psi$ (2.6$\sigma$),
consistent with the $Y(4260)$ tail, and for $\psi(4040) \to \pi^+ \pi^- J/\psi$
(3.3$\sigma$).

The hybrid interpretation of $Y(4260)$ deserves further attention.  One
consequence is a predicted decay to $D \bar D_1 +$ c.c., where $D_1$
is a P-wave $c \bar q$ pair.  Now, $D \bar D_1$ threshold is 4287 MeV/$c^2$
if we consider the lightest $D_1$ to be the state noted in Ref.\ \cite{PDG}
at 2422 MeV/$c^2$.  In this case the $Y(4260)$ would be a $D \bar D_1 +$ c.c.
{\it bound state}.  It would decay to $D \pi \bar D^*$, where the $D$ and $\pi$
are not in a $D^*$. The dip in $R_{e^+ e^-}$ lies just below $D \pi \bar
D^*$ threshold, which may be the first S-wave meson pair accessible in
$c \bar c$ fragmentation \cite{Close:2005iz}.

\subsection{Charmonium: updated}

\begin{figure}
\includegraphics[width=0.98\textwidth]{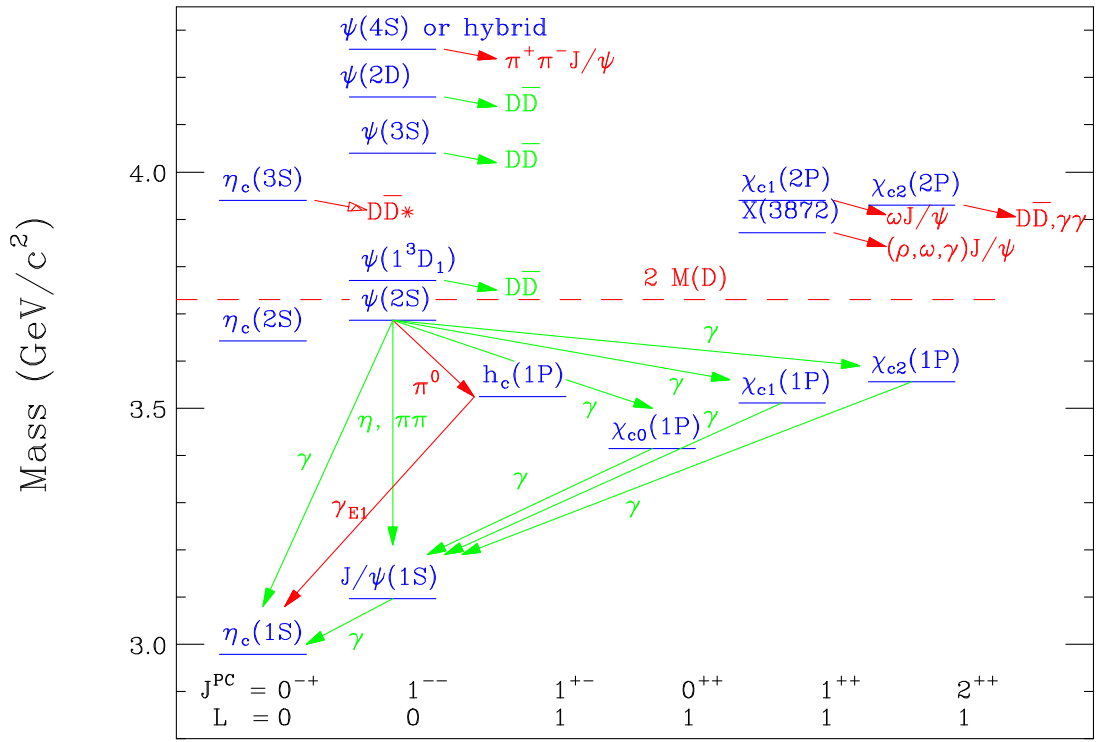}
\caption{Charmonium states including levels above charm threshold.
\label{fig:charmon}}
\end{figure}

Remarkable progress has been made in the spectroscopy of charmonium states
above charm threshold in the past few years.  Fig.\ \ref{fig:charmon}
summarizes the levels (some of whose assignments are tentative).  Even though
such states can decay to charmed pairs (with the possible exception of
$X(3872)$, which may be just below $D \bar D^*$ threshold), other decay modes
are being seen.  I have not had time to discuss much other interesting work
by BES and CLEO on exclusive decays of the $\chi_{cJ}$ and $\psi(2S)$ states,
including studies of strong-electromagnetic interference in $\psi(2S)$ decays.

\begin{figure}
\includegraphics[height=0.4\textheight]{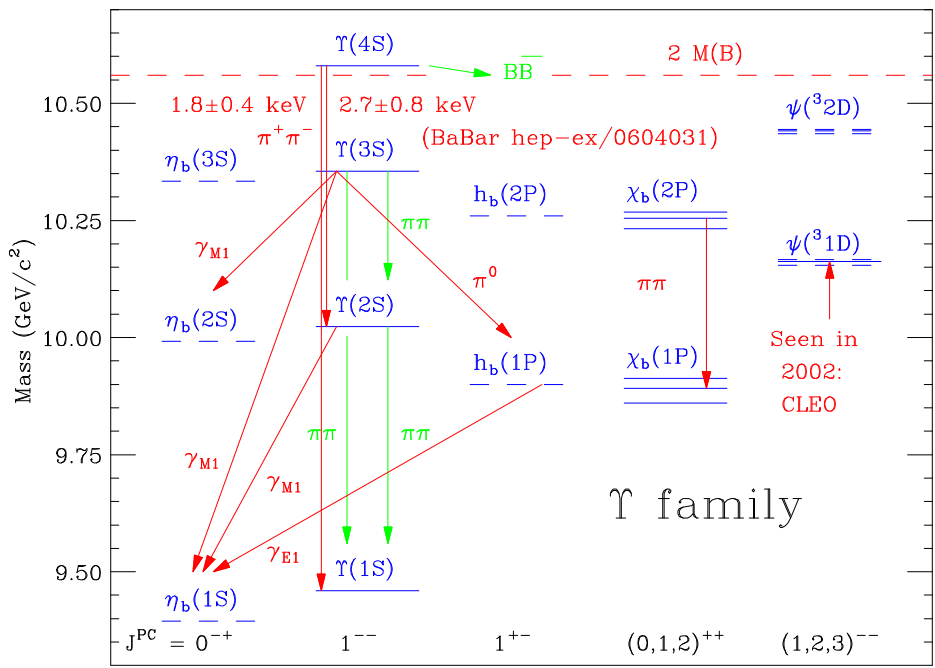}
\caption{$b \bar b$ levels and some decays.  Electric dipole (E1) transitions
$S \leftrightarrow P \leftrightarrow D$ are not shown.
\label{fig:ups}}
\end{figure}

\section{THE $\Upsilon$ FAMILY (BOTTOMONIUM)}

Some properties and decays of the $\Upsilon$ ($b \bar b$) levels are summarized
in Fig. \ref{fig:ups}.  Masses are in agreement with unquenched lattice QCD
calculations, a triumph of theory \cite{Lepage}.  Direct photons have been
observed in 1S, 2S, and 3S decays, implying estimates of the strong
fine-structure constant consistent with others \cite{Besson:2005jv}.  The
transitions $\chi_b(2P) \to \pi \pi \chi_b(1P)$ have been seen
\cite{Cawlfield:2005ra,Tati}.  In addition to the $\Upsilon(4S) \to \pi^+ \pi^-
\Upsilon(1S,2S)$ transitions noted in Fig.\ \ref{fig:ups} \cite{BaUps}, Belle
has seen $\Upsilon(4S) \to \pi^+ \pi^- \Upsilon(1S)$, with a branching ratio
${\cal B} = (1.1 \pm 0.2 \pm 0.4) \times 10^{-4}$ \cite{BeUps}.

\subsection{Remeasurement of $\Upsilon(nS)$ properties}
\vskip -0.1in

New values of ${\cal B}[\Upsilon(1S,2S,3S) \to \mu^+ \mu^-] = (2.39 \pm 0.02
\pm 0.07,~2.03\pm0.03\pm0.08,~2.39\pm0.07\pm0.10)\%$ \cite{Adams:2004xa},
when combined with new measurements $\Gamma_{ee}(1S,2S,3S) = (1.354\pm0.004
\pm0.020,~0.619\pm0.004,\pm0.010,~0.446\pm0.004\pm0.007)$ keV imply total
widths $\Gamma_{\rm tot}(1S,2S,3S) = (54.4\pm0.2\pm0.8\pm1.6,~30.5\pm0.2\pm
0.5\pm1.3,~18.6\pm0.2\pm0.3\pm0.9)$ keV.  The values of $\Gamma_{\rm tot}
(2S,3S)$ are significantly below world averages \cite{PDG}, which will lead to
changes in comparisons of predicted and observed transition rates.  As one
example, the study of $\Upsilon(2S,3S) \to \gamma X$ decays
\cite{Artuso:2004fp} has provided new branching ratios for E1 transitions
to $\chi_{bJ}(1P),~\chi'_{bJ}(2P)$ states.  These may be combinedwith the
new total widths to obtain updated partial decay widths [line (a) in
Table \ref{tab:E1}], which may be compared with one set of non-relativistic
predictions \cite{KR} [line (b)].  The suppression of transitions to $J=0$
states by 10--20\% with respect to non-relativistic expectations agrees
with relativistic predictions \cite{rel}.  The partial width for $\Upsilon(3S)
\to \gamma 1^3P_0$ is found to be $56 \pm 20$ eV, about eight times the
highly-suppressed value predicted in Ref.\ \cite{KR}.  That prediction is
very sensitive to details of wave functions; the discrepancy indicates
the importance of relativistic distortions.

\begin{table}
\caption{Comparison of observed (a) and predicted (b) partial widths
for $2S \to 1 P_J$ and $3S \to 2 P_J$ transitions in $b \bar b$ systems.
\label{tab:E1}}
\begin{tabular}{|c|c c c|c c c|} \hline
 & \multicolumn{3}{c|}{$\Gamma$ (keV), $2S \to 1P_J$ transitions}
 & \multicolumn{3}{c|}{$\Gamma$ (keV), $3S \to 2P_J$ transitions} \\
 & $J=0$ & $J=1$ & $J=2$ & $J=0$ & $J=1$ & $J=2$ \\ \hline
(a) & 1.14$\pm$0.16 & 2.11$\pm$0.16 & 2.21$\pm$0.16 &
 1.26$\pm$0.14 & 2.71$\pm$0.20 & 2.95$\pm$0.21 \\
(b) &     1.39       &     2.18      &     2.14      &
     1.65      &     2.52       &     2.78       \\ \hline
\end{tabular}
\end{table}

\subsection{$b \bar b$ spin singlets}
\vskip -0.1in

Decays of the $\Upsilon(1S,2S,3S)$ states are potential sources of information
on $b \bar b$ spin-singlets, but none has been seen yet.  One expects
1S, 2S, and 3S hyperfine splittings to be approximately 60, 30, 20 MeV/$c^2$,
respectively \cite{Godfrey:2001eb}.  The lowest P-wave singlet state (``$h_b$'')
is expected to be near $\langle M(1^3P_J) \rangle \simeq 9900$ MeV/$c^2$
\cite{Godfrey:2002rp}.

Several searches have been performed or are under way in 1S, 2S, and 3S CLEO
data.  One can search for the allowed M1 transition in $\Upsilon(1S) \to \gamma
\eta_b(1S)$ by reconstructing exclusive final states in $\eta_b(1S)$ decays
and dispensing with the soft photon, which is likely to be swallowed up in
background.  Final states are likely to be of high multiplicity.

One can search for higher-energy but suppressed M1 photons in $\Upsilon(n'S)
\to \gamma \eta_b(nS)~(n \ne n')$ decays.  These searches already exclude many
models. The strongest upper limit obtained is for $n'=3$, $n=1$: ${\cal B} \le
4.3 \times 10^{-4}$ (90\% c.l.).  $\eta_b$ searches using sequential processes
$\Upsilon(3S) \to \pi^0 h_b(1^1P_1) \to \pi^0 \gamma \eta_b(1S)$ and
$\Upsilon(3S) \to \gamma \chi'_{b0} \to \gamma \eta \eta_b(1S)$ (the latter
suggested in Ref.\ \cite{Voloshin:2004hs}) are being conducted but there are no
results yet.  Additional searches for $h_b$ involve the transition
$\Upsilon(3S) \to \pi^+ \pi^- h_b$ [for which a typical experimental upper
bound based on earlier CLEO data \cite{Brock:1990pj} is
${\cal O}(10^{-3}$)], with a possible $h_b \to \gamma \eta_b$ transition
expected to have a 40\% branching ratio \cite{Godfrey:2002rp}.

\section{FUTURE PROSPECTS}

Two main sources of information on hadron spectroscopy in the past few years
have been BES-II and CLEO. BES-II has ceased operation to make way for BES-III.
CLEO's original goals of 3 fb$^{-1}$ at $\psi(3770)$, 3 fb$^{-1}$ above $D_s$
pair threshold, and $10^9$ $J/\psi$ now appear unrealistic in light of
attainable CESR luminosity.  Consequently, it was agreed to focus CLEO on 3770
and 4170 MeV, split roughly equally, yielding about 750 pb$^{-1}$ at each
energy if current luminosity projections hold.  The determination of $f_D$,
$f_{D_s}$, and form factors for semileptonic $D$ and $D_s$ decays will provide
incisive tests for lattice gauge theories and measure CKM factors $V_{cd}$ and
$V_{cs}$ with unprecedented precision.  A sample of 30 million $\psi(2S)$
(about 10 times the current number) is planned to be taken, with at least 10
million this summer.  Some flexibility to explore new phenomena will be
maintained.  CLEO-c running will end at the end of March 2008; BES-III and
and PANDA will carry the torch thereafter.

Belle has taken 3 fb$^{-1}$ of data at $\Upsilon(3S)$; it is anyone's guess what
they will find with such a fine sample.  For comparison, CLEO has (1.1,1.2,1.2)
fb$^{-1}$ at (1S,2S,3S).  Both BaBar and Belle have shown interest in hadron
spectroscopy and are well-positioned to study it.  There have been significant
contributions from CDF and D0 as well, and we look forward to more.

\section{SUMMARY}

Hadron spectroscopy is providing both long-awaited states like $h_c$ (whose
mass and production rate confirm theories of quark confinement and
isospin-violating $\pi^0$-emission transitions) and surprises like low-lying
P-wave $D_s$ mesons, X(3872), X(3940), Y(3940), Z(3940) and Y(4260).  Decays of
$\psi''(3770)$ shed light on its nature as a $1^3D_1$ $c \bar c$ state with
a small S-wave admixture.

Upon reflection, some properties of the new hadron states may be less
surprising but we are continuing to learn about properties of QCD in the
strong-coupling regime.  There is evidence for molecules, 3S, 2P, 4S or hybrid
charmonium, and interesting decays of states above flavor threshold.

QCD may not be the last strongly coupled theory with which we have to deal.
The mystery of electroweak symmetry breaking or the very structure of quarks
and leptons may require related techniques.  It is important to realize that
insights on hadron spectra are coming to us in general from experiments at
the frontier of intensity and detector capabilities rather than energy,
and illustrate the importance of a diverse approach to the fundamental
structure of matter.

\begin{theacknowledgments}

We owe a collective debt to R. H. Dalitz for teaching us that in order to learn
about fundamental physics such as parity violation in the weak interactions or
the existence of quarks it is often necessary to deal with phenomenological
techniques of strong-interaction physics such as ``phase space plots'' or
baryon resonance descriptions.  Let us keep Dalitz's legacy alive in our
approach to particle physics.

I am grateful to R. Faccini and other colleagues on BaBar, Belle, and CLEO for 
sharing data and for helpful discussions, and to E. van Beveren, D.  Bugg, F.
Close, J. Pel\'aez, and G. Rupp for some helpful references.  This work was
supported in part by the United States Department of Energy under Grant No.\
DE FG02 90ER40560.

\end{theacknowledgments}

\end{document}